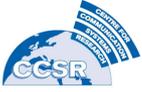

# *Saratoga*: scalable, speedy data delivery for sensor networks

Lloyd Wood

*Research Fellow, Centre for Communication Systems Research at the University of Surrey, e-mail: L.Wood@surrey.ac.uk*

**Abstract**

A networking transport protocol, named *Saratoga*, has been developed at the University of Surrey for efficient delivery of imagery from Internet-Protocol-based remote-sensing satellites. *Saratoga* is now being implemented and evaluated for use for the high-end data-delivery needs of astronomers using large, advanced, radio telescopes. These telescopes are expected to take advantage of Internet technologies. This brief paper outlines the reasons for the creation and adoption of this protocol, discusses how it differs from and complements other protocols, and summarises the worldwide collaboration that is making this development possible.

## 1. Introduction

The Disaster Monitoring Constellation (DMC), a group of low-Earth-orbiting remote-sensing satellites, is operated under a public charter. The DMC is built and run by Surrey Satellite Technology Ltd (SSTL), a successful spinoff of the University of Surrey. The satellites transfer Earth imaging data to ground while passing over a ground station. Unusually for civil satellites, the data is delivered using well-known standard terrestrial networking protocols: the Internet Protocol (IP), Frame Relay, and High-Level Data Link Control (HDLC). This approach differs from the usual custom-for-space communication standards adopted by the aerospace industry, but has proven to be both cost-effective and efficient in implementation and operation. The one piece of custom engineering needed to support the efficient IP-based delivery of sensor data is a high-speed data transfer protocol: *Saratoga*. This was designed to copy files reliably and as quickly as possible, to get the most data down during each minutes-long satellite pass over a ground station. Delivery of sensor data from the DMC satellites using IP and *Saratoga* has been relied on since 2004, and in the aftermaths of earthquakes, tsunamis, hurricanes, and other significant events.

An effort was made, in collaboration with NASA Glenn Research Center in Ohio, to work further on this novel protocol. *Saratoga*'s use allowed the first in-space tests of the 'Interplanetary Internet' to be carried out by NASA Glenn, as one of a number of experiments in taking the Internet into space with SSTL's UK-DMC satellite [1].

A public specification of the protocol allows public scrutiny and lets anyone implement *Saratoga* [2]. Defining the public specification allowed new capabilities to be added, so that *Saratoga* would be useful for future, more powerful, imaging satellites. Two major needs were identified:

- The need to support very large files, as sensors improve and communication links increase in speed. Any threshold in 'maximum filesize supported' would eventually be breached. Other file delivery protocols have filesize limits that are regularly encountered, *e.g.* a 16 MB or 32 MB file limit in the Trivial File Transfer Protocol (TFTP). The protocol had to be scalable.
- The need to support real-time continuous data delivery ('streaming') outside the file-based paradigm, so that the satellites could just 'say what they see' to a local receiving groundstation, rather than only capturing and storing a small swath of imagery for later download.

Developing an enhanced 'version one' of *Saratoga*, as an upgrade to the existing operational 'version zero' already in use by SSTL, drew interest from the astronomy community, which has similar needs. New sensor arrays on the ground, called Very Long Baseline Interferometers (VLBIs), coordinate a large number of distributed sensor inputs accurately in real time to piece together detailed views of the sky. Large planned VLBIs, such as the Square Kilometre Array (SKA), are expected to span entire continents once built. Like the DMC satellite scenario, these are also private networks with a need for high-speed data delivery – and their engineers want to leverage developments in commercial IP networking technology. *Saratoga* is now being implemented for these radio astronomy sensor networks [3].

## 2. Technical Approach

*Saratoga* builds upon the existing TCP/IP stack common to all Internet-connected computers. The Transmission Control Protocol (TCP) is widely used for file delivery, and is relied upon by the web. Fairness in use of the public Internet is a problem; if the Internet is too overloaded and congested, communications are slowed or lost. To prevent a 'tragedy of the commons,' or 'congestion collapse' of the Internet, feedback algorithms were placed between TCP senders and receivers, to detect packet loss and slow TCP's output when the network is busy. While this allows the Internet to continue to work, it means that TCP, which is adapted for shared use of public networks, is not optimal over dedicated links in private networks, where as much data as possible must be transferred as quickly as possible. TCP probes available network path capacity with slow start, and cautiously assumes that any loss due to channel errors is a deliberate packet discard by routers, indicating congestion. TCP always assuming the worst about available network capacity leads to poor delivery throughput and link underutilization. *Saratoga* does not want any form of congestion control for its environments. Instead of using TCP and its congestion control algorithms, *Saratoga* builds on the parallel User Datagram Protocol (UDP). UDP is used for real-time data delivery, *e.g.* in Voice over IP (VoIP) or gaming.

Unlike TCP, UDP does not ensure delivery; *Saratoga* adds reliable delivery to UDP. Avoiding fair, TCP-like, congestion control and a simple implementation enables *Saratoga* to run unfairly, at line speed, and output data as fast as an interface can send it. This compares well with TCP constantly probing available path capacity, overflowing, and backing off, which leads to TCP to repeatedly oscillating its throughput rate and wasting link capacity [Fig. 1]. *Saratoga* is used only over the private DMC satellite links to ground stations to deliver sensor data. Other, congestion-aware, file transfer protocols then deliver the processed imagery across the Internet to end users.

Although *Saratoga* is intended for high-speed delivery in private networks, it could unfairly congest the public Internet if used inappropriately in that environment. Showing that a protocol can be used safely in a fair 'TCP-friendly' manner is important to the Internet Engineering Task Force as a standards body. We showed that *Saratoga*'s Selective Negative Acknowledgement (SNACK) mechanism, which indicates lost packets to enable reliable delivery, can also be used for congestion control without altering the protocol [4].

By indicating and using varying-size protocol header fields, *Saratoga* can scale efficiently to handle files of sizes up to $2^{128}$ bytes (over $10^{14}$ yottabytes). That is far larger than most current filesystems, but promises to be future-proof as storage capacities and needs continue to increase. After all, twenty years ago, could you have imagined owning and using gigabytes of data in a memory stick or music player?

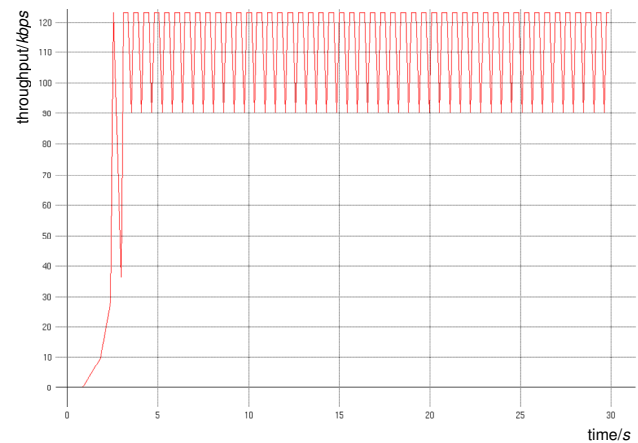

a. simulation of a single TCP flow across a reliable long-distance 128 kbps link. TCP repeatedly probes available capacity and increases its rate to above the link rate, causing the output queue to drop packets, leading to backoff once TCP notices each loss.

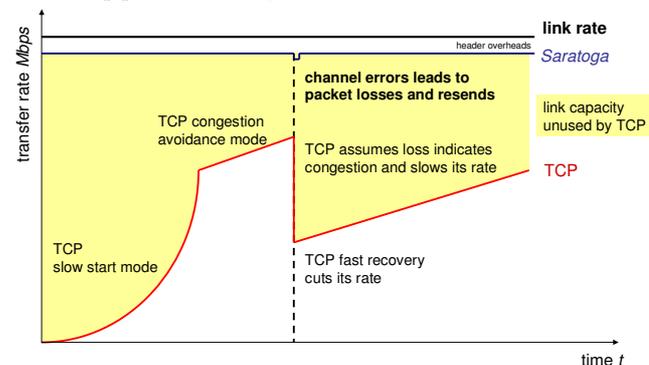

b. Identification of various TCP behaviours.

**Fig. 1 – *Saratoga* and TCP reacting to packet loss [3]**

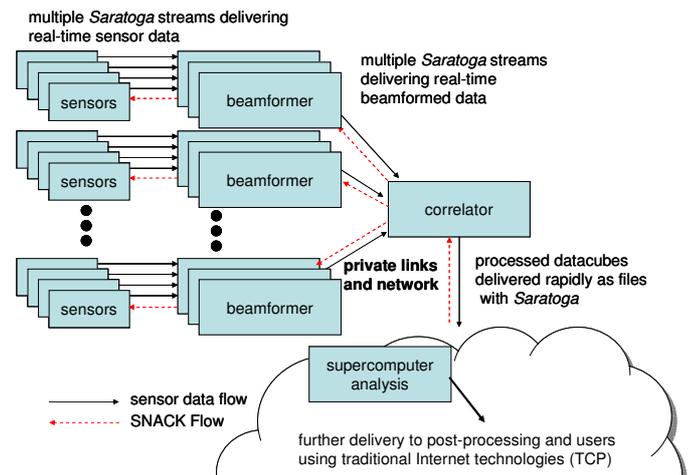

**Fig. 2 – Use of *Saratoga* in radio astronomy [3]**

### 3. Results

Wide interest in *Saratoga* bodes well for its further development and adoption. The University of Oklahoma has written a *Saratoga* simulator for research [4]. *Saratoga* is used daily by SSTL on its DMC satellites [1]. Implementations are also being developed at NASA Glenn for unmanned aviation and Earth imaging use, and by the Commonwealth Scientific and Industrial Research Organisation (CSIRO) in Sydney, Australia for use over Ethernet-based optical links in astronomy networks such as SKA [3]. That proposed use is shown [Fig. 2]. Data can be streamed from sensors at high speed by *Saratoga*, then delivered from correlators by *Saratoga* as multi-terabyte 'data cube' files for detailed analysis, before results are provided across the public Internet using other file transfer protocols. Interoperability testing between implementations is now underway.

### 4. Summary of the work, potential impact and conclusion

*Saratoga* has shown that IP and commercial networking protocols can be relied on in daily operation to deliver ever-larger amounts of mission-critical imaging data from space. We expect the same to hold true for upcoming trials of *Saratoga* for astronomy use, enabling efficient and cost-effective network engineering for radio telescopes.